\documentclass[twocolumn,amsmath,amssymb,floatfix,prb,shownopacs,footinbib,superscriptaddress]{revtex4-1}
\usepackage{graphicx,natbib}

\begin{document}

\title{Primary thermometry triad at 6 mK in mesoscopic circuits}

\author{Z. Iftikhar}
\affiliation{Centre de Nanosciences et de Nanotechnologies (C2N), CNRS, Univ Paris Sud-Universit\'e Paris-Saclay, Universit\'e Paris Diderot-Sorbonne Paris Cit\'e, 91120 Palaiseau, France}
\author{A. Anthore}
\affiliation{Centre de Nanosciences et de Nanotechnologies (C2N), CNRS, Univ Paris Sud-Universit\'e Paris-Saclay, Universit\'e Paris Diderot-Sorbonne Paris Cit\'e, 91120 Palaiseau, France}
\author{S. Jezouin}
\affiliation{Centre de Nanosciences et de Nanotechnologies (C2N), CNRS, Univ Paris Sud-Universit\'e Paris-Saclay, Universit\'e Paris Diderot-Sorbonne Paris Cit\'e, 91120 Palaiseau, France}
\author{F.D. Parmentier}
\affiliation{Centre de Nanosciences et de Nanotechnologies (C2N), CNRS, Univ Paris Sud-Universit\'e Paris-Saclay, Universit\'e Paris Diderot-Sorbonne Paris Cit\'e, 91120 Palaiseau, France}
\author{Y. Jin}
\affiliation{Centre de Nanosciences et de Nanotechnologies (C2N), CNRS, Univ Paris Sud-Universit\'e Paris-Saclay, Universit\'e Paris Diderot-Sorbonne Paris Cit\'e, 91120 Palaiseau, France}
\author{A. Cavanna}
\affiliation{Centre de Nanosciences et de Nanotechnologies (C2N), CNRS, Univ Paris Sud-Universit\'e Paris-Saclay, Universit\'e Paris Diderot-Sorbonne Paris Cit\'e, 91120 Palaiseau, France}
\author{A. Ouerghi}
\affiliation{Centre de Nanosciences et de Nanotechnologies (C2N), CNRS, Univ Paris Sud-Universit\'e Paris-Saclay, Universit\'e Paris Diderot-Sorbonne Paris Cit\'e, 91120 Palaiseau, France}
\author{U. Gennser}
\affiliation{Centre de Nanosciences et de Nanotechnologies (C2N), CNRS, Univ Paris Sud-Universit\'e Paris-Saclay, Universit\'e Paris Diderot-Sorbonne Paris Cit\'e, 91120 Palaiseau, France}
\author{F. Pierre}
\email[e-mail: ]{frederic.pierre@u-psud.fr}
\affiliation{Centre de Nanosciences et de Nanotechnologies (C2N), CNRS, Univ Paris Sud-Universit\'e Paris-Saclay, Universit\'e Paris Diderot-Sorbonne Paris Cit\'e, 91120 Palaiseau, France}

\maketitle 

{\sffamily
Quantum physics emerge and develop as temperature is reduced.
Although mesoscopic electrical circuits constitute an outstanding platform to explore quantum behavior, the challenge in cooling the electrons impedes their potential.
The strong coupling of such micrometer-scale devices with the measurement lines, combined with the weak coupling to the substrate, makes them extremely difficult to thermalize below 10~mK and imposes in-situ thermometers.
Here we demonstrate electronic quantum transport at 6~mK in micrometer-scale mesoscopic circuits.
The thermometry methods are established by the comparison of three in-situ primary thermometers, each involving a different underlying physics.
The employed combination of quantum shot noise, quantum back-action of a resistive circuit and conductance oscillations of a single-electron transistor covers a remarkably broad spectrum of mesoscopic phenomena.
The experiment, performed in vacuum using a standard cryogen-free dilution refrigerator, paves the way toward the sub-millikelvin range with additional thermalization and refrigeration techniques.
}

Advances toward lower temperatures are instrumental in the fundamental exploration of quantum phenomena.
In the context of quantum electronics, typical examples are the exploration of the correlated fractional quantum Hall physics \cite{Willett1987,Pan1999,Dolev2008,Radu2008,Deng2014}, of the quantum criticality for example with multi-channel Kondo nanostructures \cite{Potok2007,Keller2015,Iftikhar2015}, or of the quantum aspects of heat \cite{Giazotto2006,Meschke2006,Jezouin2013b,Pekola2015}.  
Although commercial dilution refrigerators readily achieve temperatures in the 5--10~mK range at the mixing chamber, the pertinent value is the temperature of the electrons within the cooled quantum circuits.
Due to microwave heating, insufficient thermal contacts and electrical noise transmitted through the measurement lines, this electronic temperature is usually well above the refrigerator base temperature. 
Consequently, only rare examples demonstrate electronic temperatures significantly below 10~mK in quantum circuits.
Moreover, the concept of temperature pervades the laws of physics, and its accurate knowledge is generally imperative whenever comparing experimental measurements with theoretical predictions; however, establishing the validity of the thermometry is particularly challenging already below 50~mK.
Because of the thermal decoupling between electrons and substrate, it requires a comparison of the electronic temperature determined in-situ, in the same device, by different methods.

\begin{figure}[!htb]
\renewcommand{\figurename}{\textbf{Figure}}
\renewcommand{\thefigure}{\textbf{\arabic{figure}}}
\center
\includegraphics[width=1\columnwidth]{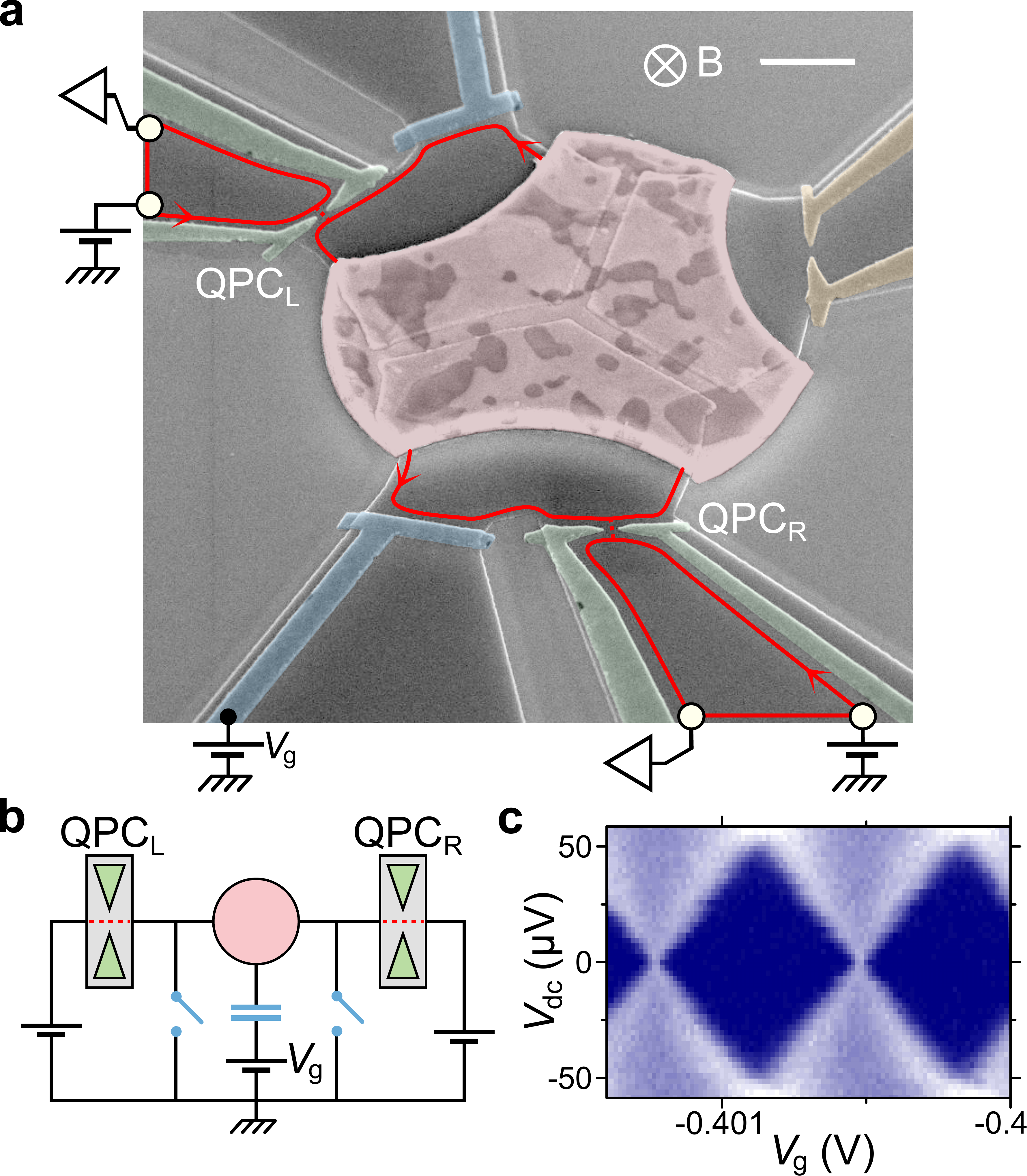}
\caption{
\footnotesize
\textbf{Cooled electrical nanostructure.} 
(\textbf{a}) Colored micrograph of the measured device. The top-right scale bar length is $1~\mathrm{\mu m}$. 
The micrometer-scale metallic island (red) is connected to 200~$\mu$m wide electrodes (represented as white circles) through two quantum point contacts (QPC, green split gates) formed in a buried 2D electron gas (2DEG, darker gray).
The lateral gates (blue) implement the switches shown in (\textbf{b}) by field effect.
The sample is immersed in a magnetic field $B$ corresponding to the integer quantum Hall regime, with the current propagating along the edge (red lines) in the direction indicated by arrows.
(\textbf{b}) Schematic electrical circuit.
Using the switches, the same device can be tuned in-situ into a voltage biased QPC, a single-electron transistor (SET), or a QPC embedded into a resistive circuit. 
(\textbf{c}) Charging energy characterization, for Coulomb blockade phenomena. 
With the device tuned into a SET, $E_C=25\pm1~\mu$eV is obtained from the height of the diamond patterns in the SET conductance (larger values shown brighter) measured versus gate ($V_{\mathrm{g}}$) and bias ($V_\mathrm{dc}$) voltages.
}
\label{fig_sample}
\end{figure}

The lowest electronic temperatures in solid-state quantum circuits were obtained in large, millimeter-scale, devices that are thereby weakly sensitive to heating through the measurement lines.
The lowest reported value of $3.7$~mK, to our knowledge, was obtained in a large array of 600 metallic islands, each $\sim100~\mu$m wide and interconnected by tunnel junctions \cite{Bradley2016}.
Comparably low temperatures, of possibly $\sim4$~mK, were inferred in 2D electron gas (2DEG) chips in the quantum Hall regime by two different teams \cite{Pan1999,Xia2000,Samkharadze2011}.
For the more broadly pertinent micrometer-scale mesoscopic circuits, the reported electronic temperatures are significantly higher.
We note the remarkably low value of 9~mK determined with current fluctuations measurements across a quantum point contact (QPC) in a 2DEG \cite{Chung2003}.
Although single-electron devices are particularly challenging, due to their high charge sensitivity, comparably low electronic temperatures, down to $\sim10$~mK, were recently demonstrated in 2DEG quantum dots \cite{Maradan2014,Potok2007,Iftikhar2015}.

Here we investigate three primary electronic thermometers, and demonstrate quantum electronic transport at 6~mK in micrometer-scale mesoscopic circuits.
For this purpose, the experiment is performed on a highly tunable 2DEG nanostructure, that can be set by field effect to different circuit configurations.
The complementary underlying physics of the thermometry methods give us access to different facets of the electronic temperature, and cover a broad spectrum of mesoscopic quantum phenomena.
Whereas quantum shot noise thermometry measures the temperature of the electronic Fermi quasiparticles, through their energy distribution \cite{Blanter2000}, the quantum back-action of a resistive circuit also probes the temperature of the electromagnetic environment \cite{Ingold1992}.
In contrast, the temperature inferred from the conductance oscillations of a single-electron transistor is very sensitive to charge fluctuations induced by non-thermal high-energy photons \cite{Glattli1997}. 
At the applied magnetic field $B=1.4$~T, we find with the quantum shot noise measured across a voltage biased quantum point contact $T_\mathrm{N}\simeq6.0\pm0.1$~mK.
From the conductance peaks across the device set to a single-electron transistor (SET) configuration, we obtain $T_\mathrm{CB}\simeq6.3\pm0.3$~mK.
From the dynamical Coulomb blockade conductance dip across two separate realizations of a QPC in series with a resistance, we find $T_\mathrm{DCBL}\simeq6\pm1$~mK and $T_\mathrm{DCBR}\simeq6.5\pm1$~mK.
The observed agreement between the three primary thermometers establishes their validity on an extended temperature range.

\begin{figure*}[!htb]
\renewcommand{\figurename}{\textbf{Figure}}
\renewcommand{\thefigure}{\textbf{\arabic{figure}}}
\center
\includegraphics[width=1\textwidth]{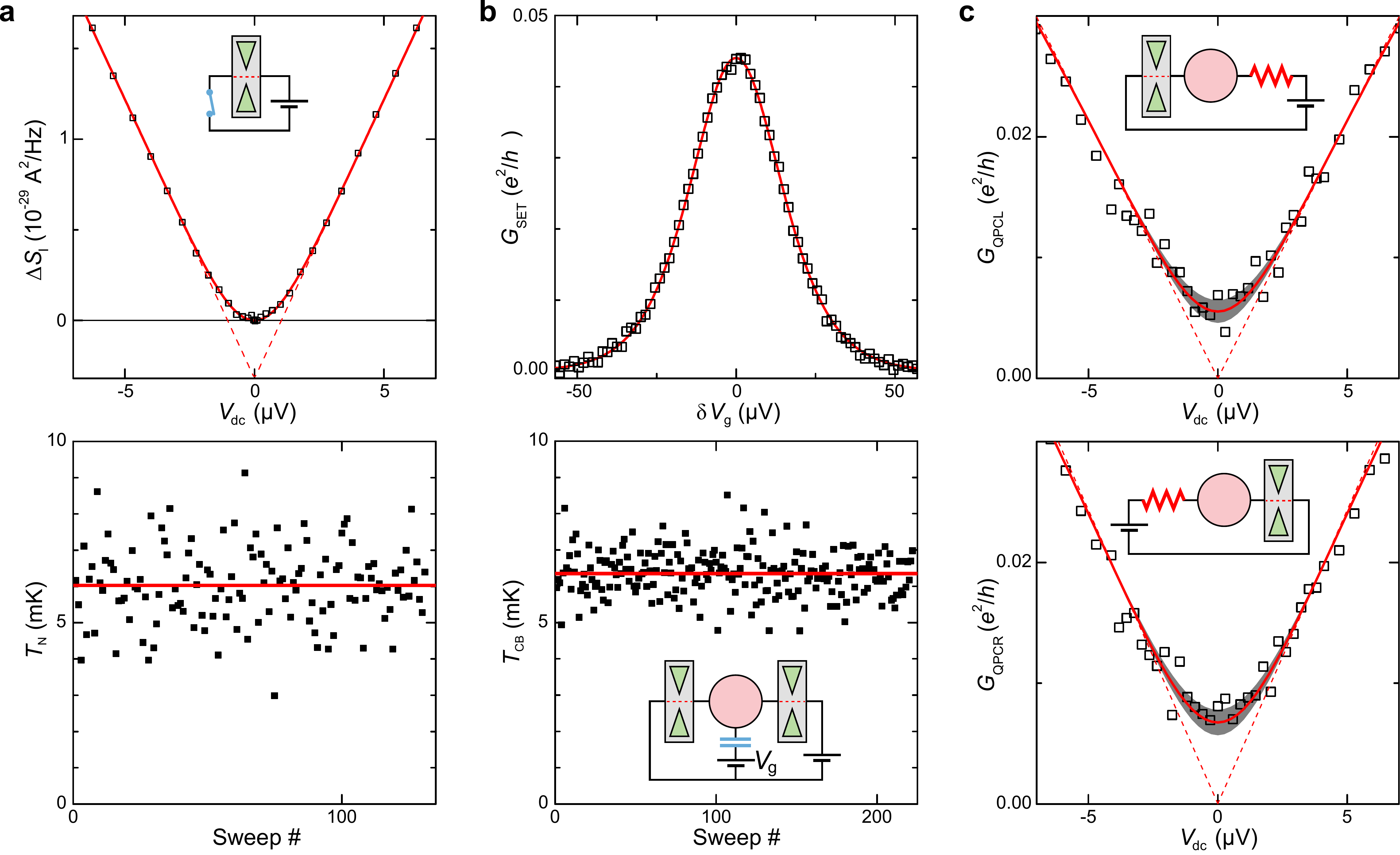}
\caption{
\footnotesize
\textbf{Primary electronic thermometry.}
(\textbf{a}) Quantum shot noise.
Symbols in the top panel represent the measured excess spectral density of the current fluctuations across a QPC biased with the dc voltage $V_\mathrm{dc}$ (see configuration schematic).
The red continuous (dashed) line is the calculated excess current fluctuations for $T_\mathrm{N}=6.0$~mK ($T_\mathrm{N}=0$, with a matching negative offset).
In the bottom panel, the different electronic temperatures $T_\mathrm{N}$ shown as symbols are each obtained by fitting a different (successive) voltage bias sweep of the quantum shot noise.
From the statistical averaging of 131 values, we find $T_\mathrm{N}\simeq6.0\pm0.1$~mK (horizontal red line) with an accuracy comparable to the provisional low temperature scale standard (PLTS-2000).
(\textbf{b}) Coulomb blockade.
Symbols in the top panel represent the measured conductance $G_\mathrm{SET}$ across the device tuned into a SET (see schematic in bottom panel) versus the gate voltage difference $\delta V_\mathrm{g}$.
The continuous line is the calculated conductance for $T_\mathrm{CB}=6.3$~mK.
The different electronic temperatures $T_\mathrm{CB}$ represented by symbols in the bottom panel are each obtained from a different gate voltage sweep $G_\mathrm{SET}(\delta V_\mathrm{g})$.
From the averaging of 222 values, we find $T_\mathrm{CB}\simeq6.3\pm0.05$~mK (horizontal red line).
Note that the accuracy on $T_\mathrm{CB}$ is limited to $\pm0.3~$mK by our uncertainty on the charging energy, $E_C=25\pm1~\mu$eV.
(\textbf{c}) Dynamical Coulomb blockade. 
The electronic temperature $T_\mathrm{DCB}$ is obtained by fitting the conductance $G_\mathrm{QPCL,R}$ of QPC$_{L,R}$ (symbols) versus voltage bias with the dynamical Coulomb blockade theory in the presence of a known series resistance $R=h/2e^2$ (see configuration schematic).
The dashed lines display the predicted suppression of the conductance at $T=0$ and $eV_\mathrm{dc}\ll E_C$, here linear in $V_\mathrm{dc}$.
We find $T_\mathrm{DCB}\simeq6\pm1~$mK ($6.5\pm1~$mK) for QPC$_L$ (QPC$_R$) from the fit shown as a continuous line in the top (bottom) panel.
The estimated uncertainty of $\pm1~$mK is displayed as a grey background.
}
\label{fig_Telec}
\end{figure*}

\vspace{\baselineskip}
{\large\noindent\textbf{Results}}\\
{\noindent\textbf{Cooled tunable mesoscopic circuit.}}
A colorized electron micrograph of the measured device is shown in Fig.~\ref{fig_sample}a, with the corresponding circuit schematic displayed Fig.~\ref{fig_sample}b.
A high-mobility 2DEG is located 105~nm below the surface of a Ga(Al)As heterojunction.
It is confined by etching within the darker grey areas delimited by bright lines, and can be tuned in-situ, by field effect, with the bias voltages applied to metallic gates deposited at the surface and capacitively coupled to the 2DEG (colorized green, yellow and blue in Fig.~\ref{fig_sample}a).
The metallic split gates at the top-left (QPC$_\mathrm{L}$) and bottom-right (QPC$_\mathrm{R}$) of Fig.~\ref{fig_sample}a (colorized green) are used to form quantum point contacts (QPCs) in the 2DEG.
Note that the split gates at the top-right of Fig.~\ref{fig_sample}a (colorized yellow) is here set to fully deplete the 2DEG underneath, thereby closing the gate, and can be ignored.
The buried 2DEG is galvanically connected, with a negligible interface resistance, to the central micrometer-sized metallic island (colorized red).
For this purpose, the metallic island was diffused into the Ga(Al)As heterojunction by thermal annealing. 
The lateral continuous gates at the surface (colorized blue) implement the equivalent of short-circuit switches in parallel with the island (blue switches in Fig.~\ref{fig_sample}b).
The experiments are performed with a magnetic field $B$ applied perpendicular to the 2DEG, which corresponds to the quantum Hall regime at integer filling factors $\nu=6$, 3 and 2 for $B=1.4$~T, $2.7$~T and $3.8$~T, respectively.
In this regime the current flows along $\nu$ chiral edge channels, represented as a single red line with the propagation direction indicated by arrows in Fig.~\ref{fig_sample}a.
Note that the quantum Hall effect is not necessary for the investigated primary thermometers (although it allows eliminating possible heating artifacts in the quantum shot noise thermometry, see Discussion).
An important device parameter is the single-electron charging energy $E_C\equiv e^2/2C$ of the central metallic island, with $C$ its overall geometrical capacitance and $e$ the elementary electron charge. 
In particular, $E_C$ sets the temperature scale extracted from Coulomb blockade thermometry.
The charging energy is most straightforwardly determined by setting the device in the SET configuration, with the short-circuit switches open (as shown in Fig.~\ref{fig_sample}a,b) and QPC$_\mathrm{L,R}$ tuned to tunnel contacts.
The SET conductance is plotted in Fig.~\ref{fig_sample}c (higher values shown brighter) versus the capacitively coupled gate voltage $V_\mathrm{g}$ and the applied drain-source dc voltage $V_\mathrm{dc}$. 
The charging energy is directly related to the periodic `Coulomb diamond' patterns in Fig.~\ref{fig_sample}c: $E_C=|eV_\mathrm{dc}^\mathrm{max}|/2\simeq25\pm1~\mu$eV, with $V_\mathrm{dc}^\mathrm{max}$ the diamonds' maximum dc voltage.

\vspace{\baselineskip}
{\noindent\textbf{Electronic current fluctuations.}}
The current across a voltage biased quantum coherent conductor fluctuates due to the thermal agitation (the Johnson-Nyquist noise) and the granularity of charge transfers (the shot noise) \cite{Blanter2000}.
These fluctuations give information on the charge of the carriers, e.g. in the fractional quantum Hall regimes \cite{Saminadayar1997,dePicciotto1997,Chung2003,Dolev2008}, as well as on the statistics of the charge transfers \cite{Reznikov1995,Kumar1996,Henny1999}, and also provide a very robust primary thermometer for the electronic temperature \cite{Spietz2003}.

We have measured the current fluctuations across the device tuned into a voltage biased QPC (schematic shown in top panel of Fig.~\ref{fig_Telec}a, see Supplementary Note 2 for details on the current fluctuations measurement setup).
For this purpose, the right short-circuit switch in Fig.~\ref{fig_sample}b was effectively closed, by applying $V_\mathrm{g}=0$ to the continuous gate adjacent to QPC$_R$.
Thereby, the 2DEG is not depleted and the edge current flows underneath the gate without back-scattering, implementing an ideal closed switch (see Supplementary Fig.~1).

The dependence with bias voltage $V_\mathrm{dc}$ of the current fluctuations' spectral density, $S_\mathrm{I}(V_\mathrm{dc})$, is directly related to the electrons' energy distribution (including in out-of-equilibrium situations \cite{Pistolesi2004,Gabelli2013}). 
For a short quantum conductor, the excess spectral density $\Delta S_\mathrm{I}(V_\mathrm{dc})\equiv S_\mathrm{I}(V_\mathrm{dc})-S_\mathrm{I}(0)$ can be calculated in the standard framework of the scattering approach \cite{Landauer1975,Anderson1980,Buttiker1986}.
It reads \cite{Blanter2000}:
\begin{equation}
\Delta S_\mathrm{I}=\frac{2e^2}{h}\sum\limits_n\tau_n(1-\tau_n)\times\left[eV_\mathrm{dc}\coth\left(\frac{eV_\mathrm{dc}}{2k_BT}\right)-2k_BT\right],
\label{eqSI}
\end{equation}
where the quantum conductor is described as a set of independent conduction channels, indexed by the label $n$, each characterized by a transmission probability $\tau_n$, and with $k_B$ ($h$) the Boltzmann (Planck) constant.
Note that the noise added by the amplification chain is canceled out by considering the excess spectral density $\Delta S_\mathrm{I}$.
Importantly, the product between the gain of the amplification chain and $\sum\tau_n(1-\tau_n)$ is given by the temperature-independent linear slope predicted at $|eV_\mathrm{dc}|\gg k_BT$.
Fitting the raw spectral density data based on Eq.~\ref{eqSI} therefore allows a self-calibrated determination of the electronic temperature, without requiring the knowledge of $\{\tau_n\}$ or of the amplification gain (see Supplementary Note 2 for further details).

The symbols in the top panel of Fig.~\ref{fig_Telec}a display the excess current spectral density measured at $B=1.4$~T versus the dc bias voltage applied across the QPC, which is tuned into the advantageous configuration of a single half-transmitted conduction channel ($\tau\simeq0.55$).
Note that in order to display the current fluctuations data in A$^2$/Hz, and although it is not necessary for extracting the electronic temperature, the effective amplification chain gain is calibrated by matching the linear bias voltage increase in the raw spectral density at large $|eV_\mathrm{dc}|\gg k_BT$ with the prediction of Eq.~\ref{eqSI} for the measured $\tau=0.55$.
The continuous (dashed) line shows $\Delta S_\mathrm{I}$ calculated using Eq.~\ref{eqSI} with $\tau=0.55$ and $T=6.0$~mK ($T=0$, with a negative vertical offset to match the $T=6.0$~mK calculation at $|eV_\mathrm{dc}|\gg k_BT$).
Experimentally, the main difficulty is to reach a sufficient resolution to accurately extract the electronic temperature.
To this aim, we developed a fully homemade cryogenic noise amplification scheme, based on high electron mobility transistors grown and nanostructured in the lab \cite{Liang2012,Dong2014}. 
Despite the unfavorable current-voltage conversion at $\nu=6$, because of the low quantum Hall resistance $h/6e^2\simeq4.3~\mathrm{k}\Omega$, we resolve $\Delta S_\mathrm{I}$ with an extremely high statistical precision of $\pm9\times10^{-32}\mathrm{A}^2/\mathrm{Hz}$, slightly smaller than the symbols' size.

Most directly, we have determined the electronic temperature and experimental uncertainty $T_\mathrm{N}=6.0\pm0.1$~mK from the mean value (red horizontal line in bottom panel of Fig.~\ref{fig_Telec}a) and statistical uncertainty of an ensemble of 131 values (symbols in bottom panel of Fig.~\ref{fig_Telec}a) independently obtained by separately fitting successive noise measurement sweeps $\Delta S_\mathrm{I}(V_\mathrm{dc})$.
Note that the $\Delta S_\mathrm{I}$ data shown in the top panel of Fig.~\ref{fig_Telec}a was obtained by averaging these successive sweeps (each resolved with an individual noise precision of $\pm10^{-30}\mathrm{A}^2/\mathrm{Hz}$).

\vspace{\baselineskip}
{\noindent\textbf{Coulomb blockade oscillations.}}
At low temperatures, $T\ll E_C/k_B$, the charge of a mesoscopic island connected through tunnel contacts is quantized in units of the elementary electron charge $e$.
This allows for the manipulation of single electrons in circuits, which has led to the field of `single-electronics' \cite{Ingold1992}.
Setting the device in the SET configuration (see schematic in bottom panel of Fig.~\ref{fig_Telec}b), charge quantization results in periodic peaks of the SET conductance $G_\mathrm{SET}$ when sweeping the capacitively coupled gate voltage $V_\mathrm{g}$.
In the presence of dc bias voltage, the peaks develop into periodic `Coulomb diamond' patterns as shown in Fig.~\ref{fig_sample}c.
The width of these conductance peaks at zero dc bias voltage constitute a well-known primary thermometer, frequently used in the context of mesoscopic physics.
For a metallic island, with a continuous density of states and connected through tunnel contact, the SET conductance reads \cite{Beenakker1991}:
\begin{equation}
G_\mathrm{SET}=\frac{G_\infty}{2}\frac{2E_C(\delta V_\mathrm{g}/\Delta)/ k_BT}{\sinh \left(2E_C(\delta V_\mathrm{g}/\Delta)/ k_BT\right)},
\label{eqGcb}
\end{equation}
with $G_\infty$ the classical (high temperature) conductance of the SET, $\Delta\simeq712\pm2~\mu$V the gate voltage period and $\delta V_\mathrm{g}$ the gate voltage difference to charge degeneracy.
Note that the Coulomb blockade thermometry is possible only with tunnel contacts.
In the presence of connected conduction channels with large transmission probabilities, the quantum fluctuations of the island's charge would average out Coulomb oscillations and thereby impede the Coulomb blockade thermometry (see Ref.~\citenum{Jezouin2016} for a characterization of charge quantization versus transmission probability on the same device).

The symbols in the top panel of Fig.~\ref{fig_Telec}b represent $G_\mathrm{SET}$ measured at $B=1.4$~T versus $\delta V_\mathrm{g}$. 
The continuous line shows the SET conductance calculated using Eq.~\ref{eqGcb} with $T=6.3$~mK, $G_\infty=0.088 e^2/h$, $\Delta=711~\mu$V and $E_C=25~\mu$eV.

Similarly to quantum shot noise thermometry, we determined the electronic temperature and statistical precision $T_\mathrm{CB}=6.3\pm0.05$~mK from an ensemble of 222 values (symbols in bottom panel of Fig.~\ref{fig_Telec}b) obtained by separately fitting individual sweeps of $G_\mathrm{SET}(\delta V_\mathrm{g})$.
The 222 sweeps are distributed among 14 adjacent Coulomb peaks, spreading over 10~mV in gate voltage.
We find the same electronic temperature, at experimental accuracy, for the different Coulomb peaks and also for the 15 or 16 measurements of each peak.
Note that our experimental accuracy $T_\mathrm{CB}\simeq6.3\pm0.3~$mK is limited by our resolution of the charging energy, $E_C\simeq25\pm1~\mu$eV.
The uncertainty is consequently much larger than the statistical precision.
Note also that despite a relatively low ac voltage of $0.35~\mu$V$_\mathrm{rms}$ applied to probe $G_\mathrm{SET}$, we estimate (using the master equation generalizing Eq.~\ref{eqGcb} to finite voltages \cite{Ingold1992}) that it is responsible for an effective increase of $0.1$~mK in $T_\mathrm{CB}$ (we have not corrected for this small effect).
Finally, we point out that the $G_\mathrm{SET}$ data shown in the top panel of Fig.~\ref{fig_Telec}b was obtained by averaging the 222 individual sweeps.

\vspace{\baselineskip}
{\noindent\textbf{Dynamical Coulomb blockade conductance renormalization.}}
The conductance of a quantum coherent conductor is progressively reduced upon cooling by the quantum back-action of the circuit in which it is embedded \cite{Ingold1992}.
This phenomenon, called dynamical Coulomb blockade, results from the granularity of charge transfers combined with Coulomb interactions.
It has been extensively studied and the theory is now well established in the simplest limit of a small tunnel conductor inserted into a linear circuit (see Ref.~\citenum{Ingold1992} and references therein; for recent developments beyond the tunnel limit see Refs.~\citenum{Altimiras2007,Parmentier2011,Mebrahtu2012,Jezouin2013}).

We consider here the case of a tunnel contact in series with a linear resistance $R$, as shown in the schematics of Fig.~\ref{fig_Telec}c.
In this configuration, the conductance at zero bias voltage (zero temperature) vanishes with temperature $T$ (bias voltage $V_\mathrm{dc}$) as $T^{2Re^2/h}$ (as $V_\mathrm{dc}^{2Re^2/h}$).
Similarly to quantum shot noise thermometry, the equilibrium ($V_\mathrm{dc}\ll k_BT/e$) to non-equilibrium ($V_\mathrm{dc}\gg k_BT/e$) crossover provides a primary electron thermometer.
In general, the electronic temperature can be extracted by fitting the conductance versus dc voltage with the full quantitative numerical prediction of the dynamical Coulomb blockade theory (see Ref.~\citenum{Joyez1997} for a formulation involving a single numerical integration).
Note that the extracted electronic temperature reflects equally the thermal energy distributions of the Fermi electron-quasiparticules, and of the bosonic electromagnetic modes of the quantum circuit.
The dynamical Coulomb blockade was previously used to probe the non-Fermi energy distribution of electrons driven out-of-equilibrium, in the presence of a thermalized $RC$ circuit \cite{Anthore2003}.

In the low-temperature and low-bias voltage regime ($k_BT,e|V_\mathrm{dc}|\ll E_C$), the primary dynamical Coulomb blockade thermometry reduces to the simple procedure described below.
The QPC conductance at low temperature ($T\ll E_C/k_B$) and at zero bias voltage $V_\mathrm{dc}=0$ reads \cite{Odintsov1991}:
\begin{equation}
G_\mathrm{QPC}(T)=\frac{G_\infty\pi^{\frac{3Re^2}{h}+\frac{1}{2}}\Gamma\left(1+\frac{Re^2}{h}\right)}{2\Gamma\left(1.5+\frac{Re^2}{h}\right)}\left(\frac{Re^2}{h}\frac{k_BT}{E_C} \right)^{\frac{2Re^2}{h}},
\label{eqGdcbT}
\end{equation}
where $G_\infty$ is the tunnel conductance in the absence of dynamical Coulomb blockade renormalization and $\Gamma(x)$ is the gamma function.
Extracting the temperature from the zero-bias conductance apparently requires a precise knowledge of both $G_\infty$ and the circuit parameters ($R$, $C$).
However, the necessary information is provided by the bias voltage dependence. 
In the non-equilibrium regime $k_BT\ll e|V_\mathrm{dc}|$ and at low energy compared to the single-electron charging energy $e|V_\mathrm{dc}|\ll E_C$, the QPC conductance reads \cite{Ingold1992}:
\begin{equation}
G_\mathrm{QPC}(V_\mathrm{dc})=\frac{G_\infty\left(\frac{\pi}{\gamma}\right)^\frac{2Re^2}{h}\left(\frac{2Re^2}{h}+1\right)}{\Gamma\left(2+\frac{2Re^2}{h}\right)}\left(\frac{Re^2}{h}\frac{e|V_\mathrm{dc}|}{E_C} \right)^{\frac{2Re^2}{h}},
\label{eqGdcbV}
\end{equation}
with $\gamma\simeq\exp(0.5772)$.
Consequently, the bias voltage exponent gives the series resistance $R$, and one can rewrite the zero bias voltage conductance as:
\begin{equation}
G_\mathrm{QPC}(T)=\frac{A(R)}{B(G_\infty,R,E_C)}\left(k_BT \right)^{\frac{2Re^2}{h}},
\label{eqGdcbTvsV}
\end{equation}
with $B(G_\infty,R,E_C)\equiv G_\mathrm{QPC}(V_\mathrm{dc})/|eV_\mathrm{dc}|^{2Re^2/h}$ calibrated from the conductance measured in the low-energy non-equilibrium regime, where Eq.~\ref{eqGdcbV} applies, and $A(R)$ a known function, straightforwardly obtained from Eqs.~\ref{eqGdcbT} and \ref{eqGdcbV}.

Here we determined the electronic temperature by setting one QPC in the tunnel regime ($G_\infty\sim0.1e^2/h$), while the other QPC was tuned to fully transmitting two electronic channels, thereby implementing a linear series resistance $R=h/2e^2$ (which is not renormalized by dynamical Coulomb blockade \cite{Parmentier2011,Mebrahtu2012,Jezouin2013,Mebrahtu2013}; obtained from a very broad and flat conductance plateau thanks to the quantum Hall effect \cite{Jezouin2013,Jezouin2013b}).
Symbols in the top (bottom) panels of Fig.~\ref{fig_Telec}c represent the conductance measured with the left (right) QPC in the tunnel regime, versus dc bias voltage, at $B=1.4$~T.
The continuous lines display the quantitative numerical calculations of the dynamical Coulomb blockade prediction using the separately characterized $E_C=25~\mu$eV and $R=h/2e^2$ (also corresponding to the linear bias voltage dependence), and with $G_\infty=0.123e^2/h$, $T=6$~mK for the top panel ($G_\infty=0.139e^2/h$, $T=6.5$~mK for the bottom panel).
The grey areas represent a temperature uncertainty of $\pm1$~mK. 
The dashed lines are the $T=0$ predictions of Eq.~\ref{eqGdcbV} for the same device parameters.
Note that for the present circuit implementation $A(R=h/2e^2)\simeq0.40$ and the non-equilibrium conductance increases linearly with bias voltage, as can be directly verified on the conductance data.
See Supplementary Fig.~2 for a comparison between the numerically calculated dynamical Coulomb blockade predictions and the data up to larger bias voltages.

\begin{figure}[!htb]
\renewcommand{\figurename}{\textbf{Figure}}
\renewcommand{\thefigure}{\textbf{\arabic{figure}}}
\center
\includegraphics[width=1\columnwidth]{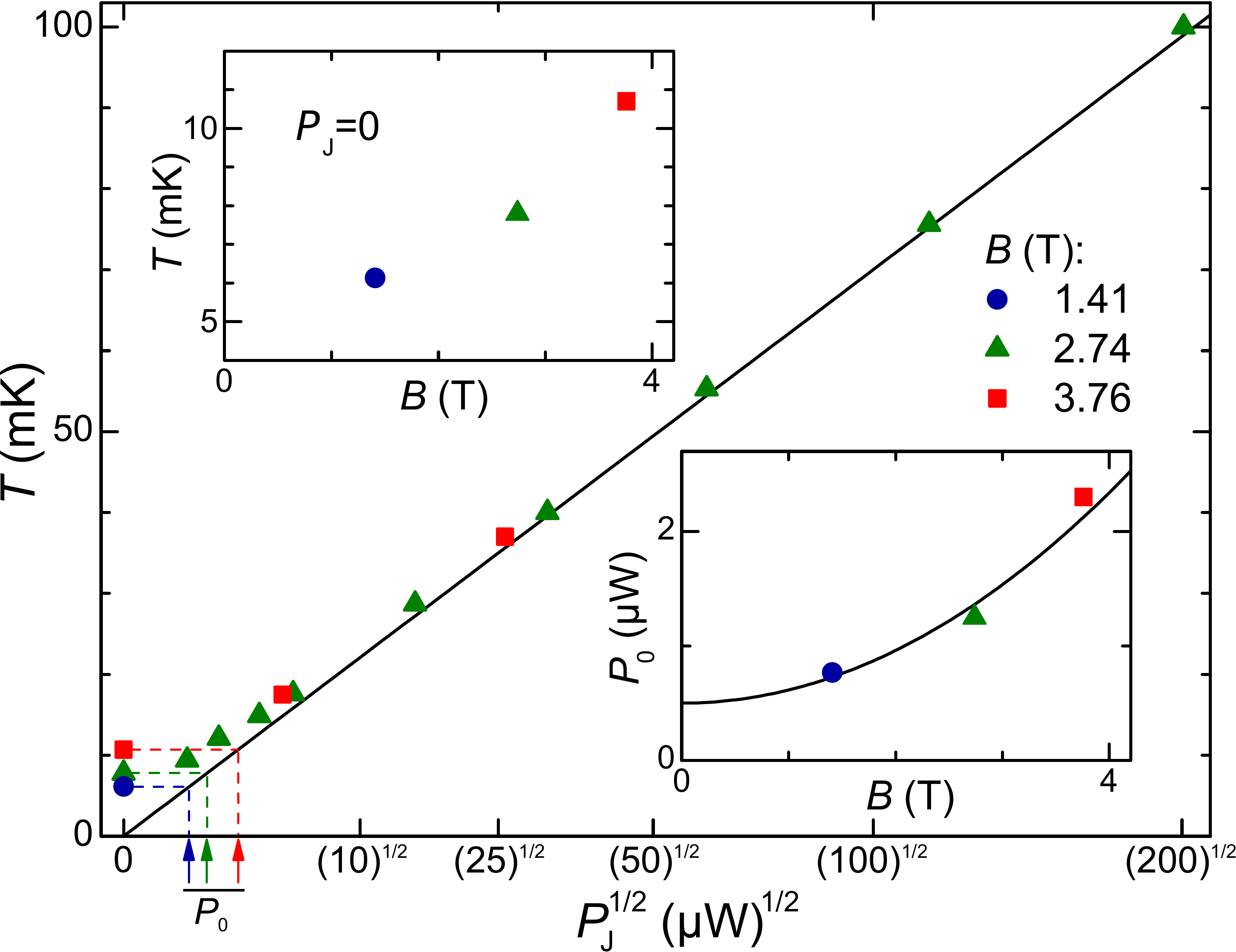}
\caption{
\footnotesize
\textbf{Temperature vs magnetic field and Joule heating.} 
The measured electronic temperature (symbols) is represented versus the square root of the Joule power $P_\mathrm{J}$ dissipated on the mixing chamber of the dilution refrigerator, for three different values of the perpendicular magnetic field $B$.
The continuous line displays $T=7\sqrt{P_\mathrm{J}/1~\mathrm{W}}$~K.
Top left inset: The temperature at $P_\mathrm{J}=0$ (symbols) increases with the applied magnetic field $B$.
Bottom right inset: The intrinsic dissipated power $P_0$, estimated assuming $T\propto\sqrt{P_0+P_\mathrm{J}}$ (arrows in main panel), is plotted as symbols versus magnetic field.
The continuous line displays $P_0=0.5+0.1(B/1~\mathrm{T})^2~\mu\mathrm{W}$.
}
\label{fig_T(B,P)}
\end{figure}

\vspace{\baselineskip}
{\noindent\textbf{Electronic temperature vs experimental conditions.}}
Information on the limiting factors toward lower electronic temperatures $T$ in our cryogen-free dilution refrigerator are obtained by measuring $T$ for different magnetic fields $B$ and for different additional Joule powers $P_\mathrm{J}$ dissipated directly on the mixing chamber plate.
Note that $T$ is here obtained from quantum shot noise thermometry up to 35~mK, and from the identical but faster readings of our standard RuO$_2$ thermometer at higher temperatures.
Each set of symbols in Fig.~\ref{fig_T(B,P)} corresponds to a different applied $B\in \{1.41,2.74,3.76\}~$T.
For $P_\mathrm{J}\gtrsim5~\mu$W, we observe the usual quadratic dependence with temperature ($T^2\propto P_\mathrm{J}$), independently of the applied $B$. 
However, we find that the electronic temperature at zero Joule power is higher for larger magnetic fields (top left inset).
Assuming that the observed relationship $T=7\sqrt{P_\mathrm{J}/1~\mathrm{W}}~$K (straight black line in main panel) holds at all temperatures when substituting the additional Joule power by the full dissipated power $P=P_0+P_\mathrm{J}$, we extract the refrigerator dissipated power $P_0$ versus magnetic field.
The corresponding $P_0$ values are shown as symbols in the bottom right inset.
We find that the increase of $P_0$ with $B$ is compatible with a quadratic magnetic field dependence (continuous black line: $P_0=0.5+0.1 (B/1~\mathrm{T})^2~\mu$W), which is a typical signature of eddy current dissipation.

\begin{figure}[!htb]
\renewcommand{\figurename}{\textbf{Figure}}
\renewcommand{\thefigure}{\textbf{\arabic{figure}}}
\center
\includegraphics[width=1\columnwidth]{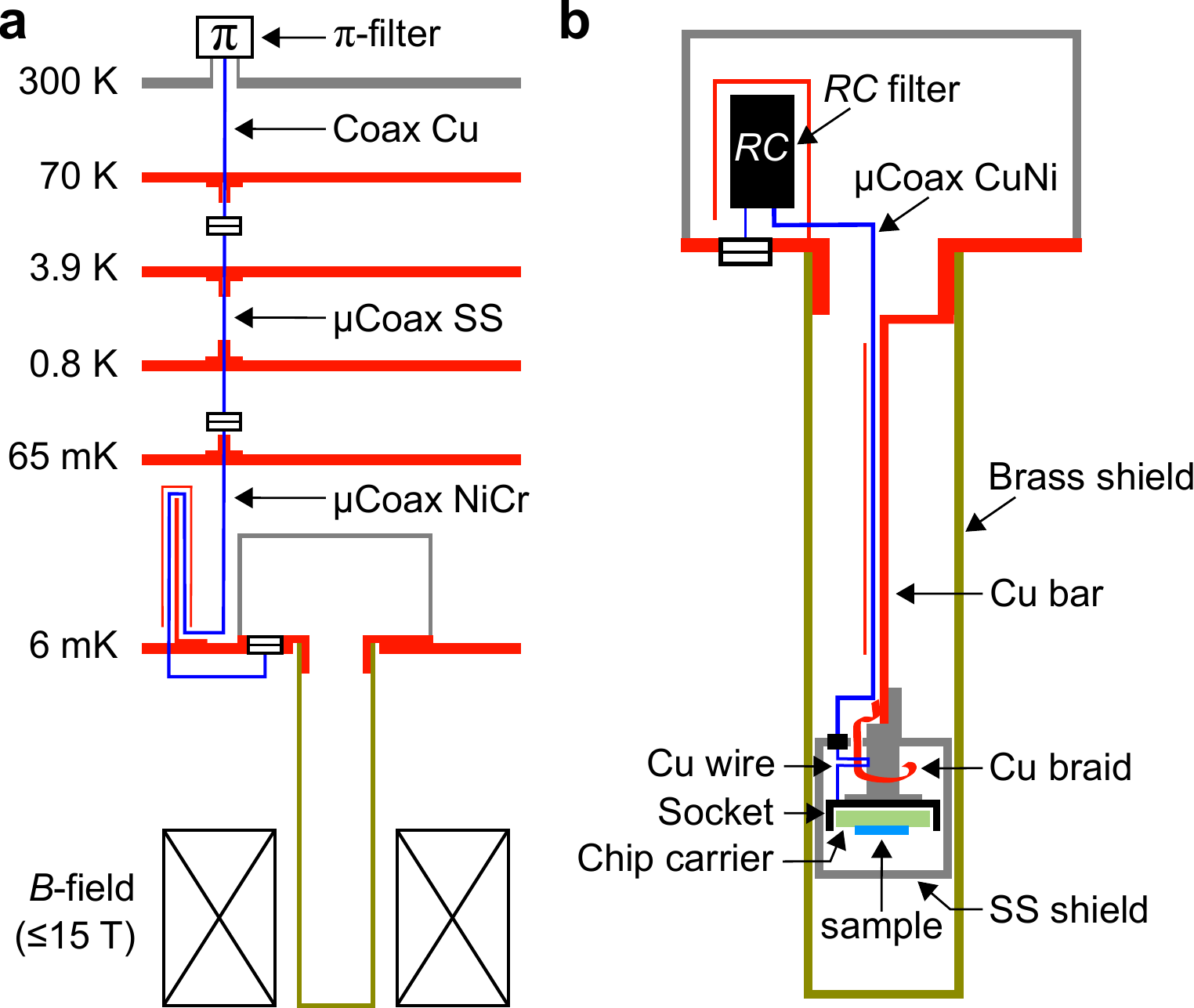}
\caption{
\footnotesize
\textbf{Experimental setup.} 
(\textbf{a}) Diagrammatic representation of the electrical lines between room temperature ($\sim300~$K) and the top-loaded sample holder at base temperature.
(\textbf{b}) Schematic representation of the top-loaded sample holder.
}
\label{fig_schem}
\end{figure}

\vspace{\baselineskip}
{\large\noindent\textbf{Discussion}}\\
A 6~mK electronic temperature was obtained in micrometer-scale quantum circuits using a medium-sized cryogen-free dilution refrigerator (Oxford instruments Triton, with $200~\mu$W of cooling power at 100~mK), with the sample in vacuum and in the presence of a $1.4$~T magnetic field.
At larger magnetic fields $B$, we observe a temperature increase that corresponds to an additional dissipated power quadratic in $B$, as typically expected for eddy currents.
In our cryogen-free refrigerator, the underlying vibrations originate from the pulse tube.

The sample environment and wiring shown Fig.~\ref{fig_schem} offers a proven guideline to ultra-low electronic temperatures with an all-purpose setup, including 35 measurement lines and a top-loaded sample holder. 
Although additional details are provided in the Supplementary Note 1, we here briefly point out several key ingredients.
The sample is strongly protected from spurious high-energy photons, by two shields at base temperature.
The most important thermal anchoring of the measurement lines at base temperature is performed by dipping insulated copper wires into silver epoxy very close to the sample, inside the inner stainless steel (SS) shield.
The measurement lines are all individually shielded in a coaxial cable geometry (except for the above mentioned copper wires and for a short distance inside the shielded sample holder, between the input connector and the $RC$ filters).
The electrical lines high frequency filtering and initial thermalization to the mixing chamber plate are performed with homemade resistive microcoaxes ($\mu$Coax NiCr in Fig.~\ref{fig_schem}a) \cite{Glattli1997}.
Because the electrical noise integrated over the full bandwidth needs to be smaller than a fraction of $\mu$V, we only keep the bandwidth used for the measurements with personalized $RC$ filters directly located inside the sample holder.
This is most particularly important with a cryogen-free dilution refrigerator in the presence of a magnetic field, due to the electrical noise induced by vibrations.

We now compare the three investigated primary electronic thermometers.

Quantum shot noise thermometry stands out as the most robust and straightforward approach.
It is based on simple physics, directly probes the temperature of the electrons through their energy distribution \cite{Blanter2000}, and does not require a separate calibration of the noise measurement setup.
The main possible artifact is local heating induced by the dissipated Joule power at finite dc bias.
Such a heating typically scales linearly with $V_\mathrm{dc}$ \cite{Smith1986,Wellstood1994}.
It is therefore difficult to distinguish from a slight increase in the shot noise \cite{Kumar1996,Henny1999}.
The present implementation in the quantum Hall regime, however, provides a strong protection against heating artifacts, thanks to the spatial separation between incoming and outgoing currents.
Although it is not necessary to determine the factor $\sum \tau_n(1-\tau_n)$ for the voltage biased quantum conductor, it is important to make sure that it does not depend on $V_\mathrm{dc}$.
For a single channel quantum conductor, the dependence of $\tau(1-\tau)$ with voltage bias is minimized at $\tau\sim0.5$, and $\tau(V_\mathrm{dc})$ can be monitored simultaneously with the noise measurements.
The main challenge with quantum shot noise thermometry is in the sensitivity of the noise measurement setup; however the associated temperature uncertainty can be statistically quantified.
Note that the achieved resolution of $6.0\pm0.1$~mK is comparable to the accuracy of the provisional low temperature scale standard (PLTS-2000) \cite{Rusby2002}.

Coulomb blockade thermometry is also very straightforward and has the advantage of being less demanding on the measurement sensitivity.
It is consequently widespread in the field of mesoscopic physics.
However, the extracted temperature is easily/often artificially increased by charge fluctuations in the device vicinity, or by the electrical noise on the capacitively coupled gates.
Such an artifact could be detected as a gate voltage dependent increase in the noise level, proportional to $\partial G_\mathrm{SET}/\partial V_\mathrm{g}$, provided that a significant part of the charge fluctuations are within the measurement bandwidth.
In general, the corresponding temperature increase is difficult to establish, except by comparing with another electronic thermometer.
Here, the agreement obtained with both the quantum shot noise and dynamical Coulomb blockade thermometers demonstrates a negligible artificial increase of the electronic temperature.

Dynamical Coulomb blockade thermometry can be difficult to use in general, if the surrounding circuit is not known a priori at the relevant GHz frequencies.
As in the case of quantum shot noise, a possible artifact is heating at finite dc bias.
This can be minimized by using a tunnel contact of very large impedance compared to the circuit.
In contrast to quantum shot noise, the quantum Hall regime does not provide a protection against heating (in the central metallic island, for the dynamical Coulomb blockade experimental configurations).
However, the very large renormalized tunnel resistance, $~100$ larger than the series resistance, ascertains negligible heating effects.
Moreover, the dynamical Coulomb blockade thermometry is here particularly straightforward to implement because of the precise knowledge of the circuit.

With the consistent temperatures obtained by three primary thermometers, each relying on different physical mechanisms, we firmly established electronic thermometry standards in the regime of ultra-low temperatures.
The achievement of 6~mK electronic temperature, with the mesoscopic circuit in vacuum and using a medium-sized dilution refrigerator, provides a platform for further reduction of the temperature, using additional thermalization and cooling techniques \cite{Clark2010,Samkharadze2011,Casparis2012}, towards the sub-millikelvin range.

\vspace{\baselineskip}
{\large\noindent\textbf{Methods}}\\
{\footnotesize
{\noindent\textbf{Sample.}}
The sample was nanostructured by standard e-beam lithography in a Ga(Al)As 2DEG of density $2.5\times10^{11}~\mathrm{cm}^{-2}$ and mobility $10^6~\mathrm{cm}^2\mathrm{V}^{-1}\mathrm{s}^{-1}$.
The AuGeNi metallic island was diffused by thermal annealing into the semiconductor heterojunction to make an electrical contact of negligible resistance with the 2DEG (see Methods in Ref.~\citenum{Iftikhar2015} for the electrical characterization of the contact in the same sample).

\vspace{\baselineskip}
{\noindent\textbf{Measurement techniques.}}
The differential conductance measurements were performed using standard lock-in techniques at frequencies below 200~Hz and using rms ac excitation voltages smaller than $k_\mathrm{B}T/e$.
The sample was current biased by a voltage source in series with a $100~$M$\Omega$ polarization resistance at room temperature.
The applied current was converted on-chip into a voltage independent of the device configuration by taking advantage of the well-defined quantum Hall resistance to an adjacent grounded electrode ($h/\nu e^2$ at filling factor $\nu$).
Similarly, the current transmitted across (reflected from) the device was converted into a voltage with the $h/\nu e^2$ quantum Hall resistance. 
The noise measurement setup includes a home-made cryogenic preamplifier and an $L$--$C$ tank circuit of resonant frequency $0.84$~MHz; see the online Supplementary Information and also the supplementary material of Ref.~\citenum{Jezouin2013b} for a more detailed description.
}

\vspace{\baselineskip}


\begin{thebibliography}{10}
\expandafter\ifx\csname url\endcsname\relax
  \def\url#1{\texttt{#1}}\fi
\expandafter\ifx\csname urlprefix\endcsname\relax\def\urlprefix{URL }\fi
\providecommand{\bibinfo}[2]{#2}
\providecommand{\eprint}[2][]{\url{#2}}

\bibitem{Willett1987}
\bibinfo{author}{Willett, R.} \emph{et~al.}
\newblock \bibinfo{title}{Observation of an even-denominator quantum number in
  the fractional quantum hall effect}.
\newblock \emph{\bibinfo{journal}{Phys. Rev. Lett.}}
  \textbf{\bibinfo{volume}{59}}, \bibinfo{pages}{1776--1779}
  (\bibinfo{year}{1987}).

\bibitem{Pan1999}
\bibinfo{author}{Pan, W.} \emph{et~al.}
\newblock \bibinfo{title}{Exact quantization of the even-denominator fractional
  quantum hall state at
  $\mathit{\ensuremath{\nu}}\phantom{\rule{0ex}{0ex}}=\phantom{\rule{0ex}{0ex}}5/2$
  landau level filling factor}.
\newblock \emph{\bibinfo{journal}{Phys. Rev. Lett.}}
  \textbf{\bibinfo{volume}{83}}, \bibinfo{pages}{3530--3533}
  (\bibinfo{year}{1999}).

\bibitem{Dolev2008}
\bibinfo{author}{Dolev, M.}, \bibinfo{author}{Heiblum, M.},
  \bibinfo{author}{Umansky, V.}, \bibinfo{author}{Stern, A.} \&
  \bibinfo{author}{Mahalu, D.}
\newblock \bibinfo{title}{Observation of a quarter of an electron charge at the
  $\nu=5/2$ quantum hall state}.
\newblock \emph{\bibinfo{journal}{Nature}} \textbf{\bibinfo{volume}{452}},
  \bibinfo{pages}{829--834} (\bibinfo{year}{2008}).

\bibitem{Radu2008}
\bibinfo{author}{Radu, I.~P.} \emph{et~al.}
\newblock \bibinfo{title}{Quasi-particle properties from tunneling in the
  $\nu=5/2$ fractional quantum hall state}.
\newblock \emph{\bibinfo{journal}{Science}} \textbf{\bibinfo{volume}{320}},
  \bibinfo{pages}{899--902} (\bibinfo{year}{2008}).

\bibitem{Deng2014}
\bibinfo{author}{Deng, N.} \emph{et~al.}
\newblock \bibinfo{title}{$\ensuremath{\nu}=5/2$ fractional quantum hall state
  in the presence of alloy disorder}.
\newblock \emph{\bibinfo{journal}{Phys. Rev. Lett.}}
  \textbf{\bibinfo{volume}{112}}, \bibinfo{pages}{116804}
  (\bibinfo{year}{2014}).

\bibitem{Potok2007}
\bibinfo{author}{Potok, R.~M.}, \bibinfo{author}{Rau, I.~G.},
  \bibinfo{author}{Shtrikman, H.}, \bibinfo{author}{Oreg, Y.} \&
  \bibinfo{author}{Goldhaber-Gordon, D.}
\newblock \bibinfo{title}{Observation of the two-channel kondo effect}.
\newblock \emph{\bibinfo{journal}{Nature}} \textbf{\bibinfo{volume}{446}},
  \bibinfo{pages}{167--171} (\bibinfo{year}{2007}).

\bibitem{Keller2015}
\bibinfo{author}{Keller, A.~J.} \emph{et~al.}
\newblock \bibinfo{title}{Universal fermi liquid crossover and quantum
  criticality in a mesoscopic system}.
\newblock \emph{\bibinfo{journal}{Nature}} \textbf{\bibinfo{volume}{526}},
  \bibinfo{pages}{237--240} (\bibinfo{year}{2015}).

\bibitem{Iftikhar2015}
\bibinfo{author}{Iftikhar, Z.} \emph{et~al.}
\newblock \bibinfo{title}{Two-channel kondo effect and renormalization flow
  with macroscopic quantum charge states}.
\newblock \emph{\bibinfo{journal}{Nature}} \textbf{\bibinfo{volume}{526}},
  \bibinfo{pages}{233--236} (\bibinfo{year}{2015}).

\bibitem{Giazotto2006}
\bibinfo{author}{Giazotto, F.}, \bibinfo{author}{Heikkil\"{a}, T.},
  \bibinfo{author}{Luukanen, A.}, \bibinfo{author}{Savin, A.} \&
  \bibinfo{author}{Pekola, J.}
\newblock \bibinfo{title}{{Opportunities for mesoscopics in thermometry and
  refrigeration: Physics and applications}}.
\newblock \emph{\bibinfo{journal}{Rev. Mod. Phys.}}
  \textbf{\bibinfo{volume}{78}}, \bibinfo{pages}{217--274}
  (\bibinfo{year}{2006}).

\bibitem{Meschke2006}
\bibinfo{author}{Meschke, M.}, \bibinfo{author}{Guichard, W.} \&
  \bibinfo{author}{Pekola, J.~P.}
\newblock \bibinfo{title}{{Single-mode heat conduction by photons}}.
\newblock \emph{\bibinfo{journal}{Nature}} \textbf{\bibinfo{volume}{444}},
  \bibinfo{pages}{187--190} (\bibinfo{year}{2006}).

\bibitem{Jezouin2013b}
\bibinfo{author}{Jezouin, S.} \emph{et~al.}
\newblock \bibinfo{title}{Quantum limit of heat flow across a single electronic
  channel}.
\newblock \emph{\bibinfo{journal}{Science}} \textbf{\bibinfo{volume}{342}},
  \bibinfo{pages}{601--604} (\bibinfo{year}{2013}).

\bibitem{Pekola2015}
\bibinfo{author}{Pekola, J.~P.}
\newblock \bibinfo{title}{Towards quantum thermodynamics in electronic
  circuits}.
\newblock \emph{\bibinfo{journal}{Nat. Phys.}} \textbf{\bibinfo{volume}{11}},
  \bibinfo{pages}{118--123} (\bibinfo{year}{2015}).

\bibitem{Bradley2016}
\bibinfo{author}{{Bradley}, D.} \emph{et~al.}
\newblock \bibinfo{title}{Nanoelectronic primary thermometer below 4 mK}.
\newblock \emph{\bibinfo{journal}{Nat. Commun.}} \textbf{\bibinfo{volume}{7}},
  \bibinfo{pages}{10455} (\bibinfo{year}{2016}).

\bibitem{Xia2000}
\bibinfo{author}{Xia, J.} \emph{et~al.}
\newblock \bibinfo{title}{Ultra-low-temperature cooling of two-dimensional
  electron gas}.
\newblock \emph{\bibinfo{journal}{Physica B}} \textbf{\bibinfo{volume}{280}},
  \bibinfo{pages}{491--492} (\bibinfo{year}{2000}).

\bibitem{Samkharadze2011}
\bibinfo{author}{Samkharadze, N.} \emph{et~al.}
\newblock \bibinfo{title}{Integrated electronic transport and thermometry at
  millikelvin temperatures and in strong magnetic fields}.
\newblock \emph{\bibinfo{journal}{Rev. Sci. Instrum.}}
  \textbf{\bibinfo{volume}{82}}, \bibinfo{pages}{053902}
  (\bibinfo{year}{2011}).

\bibitem{Chung2003}
\bibinfo{author}{Chung, Y.~C.}, \bibinfo{author}{Heiblum, M.} \&
  \bibinfo{author}{Umansky, V.}
\newblock \bibinfo{title}{Scattering of bunched fractionally charged
  quasiparticles}.
\newblock \emph{\bibinfo{journal}{Phys. Rev. Lett.}}
  \textbf{\bibinfo{volume}{91}}, \bibinfo{pages}{216804}
  (\bibinfo{year}{2003}).

\bibitem{Maradan2014}
\bibinfo{author}{{Maradan}, D.} \emph{et~al.}
\newblock \bibinfo{title}{{GaAs Quantum Dot Thermometry Using Direct Transport
  and Charge Sensing}}.
\newblock \emph{\bibinfo{journal}{J. Low Temp. Phys.}}
  \textbf{\bibinfo{volume}{175}}, \bibinfo{pages}{784--798}
  (\bibinfo{year}{2014}).

\bibitem{Blanter2000}
\bibinfo{author}{Blanter, Y.~M.} \& \bibinfo{author}{B\"uttiker, M.}
\newblock \bibinfo{title}{Shot noise in mesoscopic conductors}.
\newblock \emph{\bibinfo{journal}{Phys. Rep.}} \textbf{\bibinfo{volume}{336}},
  \bibinfo{pages}{1--166} (\bibinfo{year}{2000}).

\bibitem{Ingold1992}
\bibinfo{editor}{Grabert, H.} \& \bibinfo{editor}{Devoret, M.~H.} (eds.)
  \emph{\bibinfo{title}{Single charge tunneling}} (\bibinfo{year}{1992}),
  \bibinfo{edition}{plenum, new york} edn.

\bibitem{Glattli1997}
\bibinfo{author}{Glattli, D.~C.}, \bibinfo{author}{Jacques, P.},
  \bibinfo{author}{Kumar, A.}, \bibinfo{author}{Pari, P.} \&
  \bibinfo{author}{Saminadayar, L.}
\newblock \bibinfo{title}{A noise detection scheme with 10 mK noise temperature
  resolution for semiconductor single electron tunneling devices}.
\newblock \emph{\bibinfo{journal}{J. Appl. Phys.}}
  \textbf{\bibinfo{volume}{81}}, \bibinfo{pages}{7350--7356}
  (\bibinfo{year}{1997}).

\bibitem{Saminadayar1997}
\bibinfo{author}{Saminadayar, L.}, \bibinfo{author}{Glattli, D.~C.},
  \bibinfo{author}{Jin, Y.} \& \bibinfo{author}{Etienne, B.}
\newblock \bibinfo{title}{Observation of the $\mathit{e}\mathit{/}3$
  fractionally charged laughlin quasiparticle}.
\newblock \emph{\bibinfo{journal}{Phys. Rev. Lett.}}
  \textbf{\bibinfo{volume}{79}}, \bibinfo{pages}{2526--2529}
  (\bibinfo{year}{1997}).

\bibitem{dePicciotto1997}
\bibinfo{author}{de~Picciotto, R.} \emph{et~al.}
\newblock \bibinfo{title}{Direct observation of a fractional charge}.
\newblock \emph{\bibinfo{journal}{Nature}} \textbf{\bibinfo{volume}{389}},
  \bibinfo{pages}{162--164} (\bibinfo{year}{1997}).

\bibitem{Reznikov1995}
\bibinfo{author}{Reznikov, M.}, \bibinfo{author}{Heiblum, M.},
  \bibinfo{author}{Shtrikman, H.} \& \bibinfo{author}{Mahalu, D.}
\newblock \bibinfo{title}{Temporal correlation of electrons: Suppression of
  shot noise in a ballistic quantum point contact}.
\newblock \emph{\bibinfo{journal}{Phys. Rev. Lett.}}
  \textbf{\bibinfo{volume}{75}}, \bibinfo{pages}{3340--3343}
  (\bibinfo{year}{1995}).

\bibitem{Kumar1996}
\bibinfo{author}{Kumar, A.}, \bibinfo{author}{Saminadayar, L.},
  \bibinfo{author}{Glattli, D.~C.}, \bibinfo{author}{Jin, Y.} \&
  \bibinfo{author}{Etienne, B.}
\newblock \bibinfo{title}{Experimental test of the quantum shot noise reduction
  theory}.
\newblock \emph{\bibinfo{journal}{Phys. Rev. Lett.}}
  \textbf{\bibinfo{volume}{76}}, \bibinfo{pages}{2778--2781}
  (\bibinfo{year}{1996}).

\bibitem{Henny1999}
\bibinfo{author}{Henny, M.}, \bibinfo{author}{Oberholzer, S.},
  \bibinfo{author}{Strunk, C.} \& \bibinfo{author}{Sch\"onenberger, C.}
\newblock \bibinfo{title}{1/3-shot-noise suppression in diffusive nanowires}.
\newblock \emph{\bibinfo{journal}{Phys. Rev. B}} \textbf{\bibinfo{volume}{59}},
  \bibinfo{pages}{2871--2880} (\bibinfo{year}{1999}).

\bibitem{Spietz2003}
\bibinfo{author}{Spietz, L.}, \bibinfo{author}{Lehnert, K.~W.},
  \bibinfo{author}{Siddiqi, I.} \& \bibinfo{author}{Schoelkopf, R.~J.}
\newblock \bibinfo{title}{Primary electronic thermometry using the shot noise
  of a tunnel junction}.
\newblock \emph{\bibinfo{journal}{Science}} \textbf{\bibinfo{volume}{300}},
  \bibinfo{pages}{1929--1932} (\bibinfo{year}{2003}).

\bibitem{Pistolesi2004}
\bibinfo{author}{Pistolesi, F.}, \bibinfo{author}{Bignon, G.} \&
  \bibinfo{author}{Hekking, F. W.~J.}
\newblock \bibinfo{title}{Subgap noise of a superconductor-normal-metal tunnel
  interface}.
\newblock \emph{\bibinfo{journal}{Phys. Rev. B}} \textbf{\bibinfo{volume}{69}},
  \bibinfo{pages}{214518} (\bibinfo{year}{2004}).

\bibitem{Gabelli2013}
\bibinfo{author}{Gabelli, J.} \& \bibinfo{author}{Reulet, B.}
\newblock \bibinfo{title}{Shaping a time-dependent excitation to minimize the
  shot noise in a tunnel junction}.
\newblock \emph{\bibinfo{journal}{Phys. Rev. B}} \textbf{\bibinfo{volume}{87}},
  \bibinfo{pages}{075403} (\bibinfo{year}{2013}).

\bibitem{Landauer1975}
\bibinfo{author}{Landauer, R.}
\newblock \bibinfo{title}{Residual resistivity dipoles}.
\newblock \emph{\bibinfo{journal}{Z. Phys. B}} \textbf{\bibinfo{volume}{21}},
  \bibinfo{pages}{247--254} (\bibinfo{year}{1975}).

\bibitem{Anderson1980}
\bibinfo{author}{Anderson, P.~W.}, \bibinfo{author}{Thouless, D.~J.},
  \bibinfo{author}{Abrahams, E.} \& \bibinfo{author}{Fisher, D.~S.}
\newblock \bibinfo{title}{New method for a scaling theory of localization}.
\newblock \emph{\bibinfo{journal}{Phys. Rev. B}} \textbf{\bibinfo{volume}{22}},
  \bibinfo{pages}{3519--3526} (\bibinfo{year}{1980}).

\bibitem{Buttiker1986}
\bibinfo{author}{B\"uttiker, M.}
\newblock \bibinfo{title}{Four-terminal phase-coherent conductance}.
\newblock \emph{\bibinfo{journal}{Phys. Rev. Lett.}}
  \textbf{\bibinfo{volume}{57}}, \bibinfo{pages}{1761--1764}
  (\bibinfo{year}{1986}).

\bibitem{Liang2012}
\bibinfo{author}{Liang, Y.~X.}, \bibinfo{author}{Dong, Q.},
  \bibinfo{author}{Gennser, U.}, \bibinfo{author}{Cavanna, A.} \&
  \bibinfo{author}{Jin, Y.}
\newblock \bibinfo{title}{Input noise voltage below 1 nV/Hz$^{1/2}$ at 1 kHz in
  the HEMTs at 4.2 K}.
\newblock \emph{\bibinfo{journal}{J. Low Temp. Phys.}}
  \textbf{\bibinfo{volume}{167}}, \bibinfo{pages}{632--637}
  (\bibinfo{year}{2012}).

\bibitem{Dong2014}
\bibinfo{author}{Dong, Q.} \emph{et~al.}
\newblock \bibinfo{title}{Ultra-low noise high electron mobility transistors
  for high-impedance and low-frequency deep cryogenic readout electronics}.
\newblock \emph{\bibinfo{journal}{Appl. Phys. Lett.}}
  \textbf{\bibinfo{volume}{105}}, \bibinfo{pages}{013504}
  (\bibinfo{year}{2014}).

\bibitem{Beenakker1991}
\bibinfo{author}{Beenakker, C. W.~J.}
\newblock \bibinfo{title}{Theory of coulomb-blockade oscillations in the
  conductance of a quantum dot}.
\newblock \emph{\bibinfo{journal}{Phys. Rev. B}} \textbf{\bibinfo{volume}{44}},
  \bibinfo{pages}{1646--1656} (\bibinfo{year}{1991}).

\bibitem{Jezouin2016}
\bibinfo{author}{Jezouin, S.} \emph{et~al.}
\newblock \bibinfo{title}{Controlling charge quantization with quantum
  fluctuations}.
\newblock \emph{\bibinfo{journal}{Nature}} \textbf{\bibinfo{volume}{536}},
  \bibinfo{pages}{58--62} (\bibinfo{year}{2016}).  

\bibitem{Altimiras2007}
\bibinfo{author}{Altimiras, C.}, \bibinfo{author}{Gennser, U.},
  \bibinfo{author}{Cavanna, A.}, \bibinfo{author}{Mailly, D.} \&
  \bibinfo{author}{Pierre, F.}
\newblock \bibinfo{title}{Experimental test of the dynamical coulomb blockade
  theory for short coherent conductors}.
\newblock \emph{\bibinfo{journal}{Phys. Rev. Lett.}}
  \textbf{\bibinfo{volume}{99}}, \bibinfo{pages}{256805}
  (\bibinfo{year}{2007}).

\bibitem{Parmentier2011}
\bibinfo{author}{Parmentier, F.~D.} \emph{et~al.}
\newblock \bibinfo{title}{Strong back-action of a linear circuit on a single
  electronic quantum channel}.
\newblock \emph{\bibinfo{journal}{Nat. Phys.}} \textbf{\bibinfo{volume}{7}},
  \bibinfo{pages}{935--938} (\bibinfo{year}{2011}).

\bibitem{Mebrahtu2012}
\bibinfo{author}{Mebrahtu, H.~T.} \emph{et~al.}
\newblock \bibinfo{title}{Quantum phase transition in a resonant level coupled
  to interacting leads}.
\newblock \emph{\bibinfo{journal}{Nature}} \textbf{\bibinfo{volume}{488}},
  \bibinfo{pages}{61--64} (\bibinfo{year}{2012}).

\bibitem{Jezouin2013}
\bibinfo{author}{Jezouin, S.} \emph{et~al.}
\newblock \bibinfo{title}{Tomonaga-luttinger physics in electronic quantum
  circuits}.
\newblock \emph{\bibinfo{journal}{Nat. Commun.}} \textbf{\bibinfo{volume}{4}},
  \bibinfo{pages}{1802} (\bibinfo{year}{2013}).

\bibitem{Joyez1997}
\bibinfo{author}{Joyez, P.} \& \bibinfo{author}{Esteve, D.}
\newblock \bibinfo{title}{Single-electron tunneling at high temperature}.
\newblock \emph{\bibinfo{journal}{Phys. Rev. B}} \textbf{\bibinfo{volume}{56}},
  \bibinfo{pages}{1848--1853} (\bibinfo{year}{1997}).

\bibitem{Anthore2003}
\bibinfo{author}{Anthore, A.}, \bibinfo{author}{Pierre, F.},
  \bibinfo{author}{Pothier, H.} \& \bibinfo{author}{Esteve, D.}
\newblock \bibinfo{title}{Magnetic-field-dependent quasiparticle energy
  relaxation in mesoscopic wires}.
\newblock \emph{\bibinfo{journal}{Phys. Rev. Lett.}}
  \textbf{\bibinfo{volume}{90}}, \bibinfo{pages}{076806}
  (\bibinfo{year}{2003}).

\bibitem{Odintsov1991}
\bibinfo{author}{Odintsov, A.~A.}, \bibinfo{author}{Falci, G.} \&
  \bibinfo{author}{Sch\"on, G.}
\newblock \bibinfo{title}{Single-electron tunneling in systems of small
  junctions coupled to an electromagnetic environment}.
\newblock \emph{\bibinfo{journal}{Phys. Rev. B}} \textbf{\bibinfo{volume}{44}},
  \bibinfo{pages}{13089--13092} (\bibinfo{year}{1991}).

\bibitem{Mebrahtu2013}
\bibinfo{author}{Mebrahtu, H.~T.} \emph{et~al.}
\newblock \bibinfo{title}{Observation of majorana quantum critical behaviour in
  a resonant level coupled to a dissipative environment}.
\newblock \emph{\bibinfo{journal}{Nat. Phys.}} \textbf{\bibinfo{volume}{9}},
  \bibinfo{pages}{732--737} (\bibinfo{year}{2013}).

\bibitem{Smith1986}
\bibinfo{author}{Smith, C.~G.} \& \bibinfo{author}{Wybourne, M.~N.}
\newblock \bibinfo{title}{Electric field heating of supported and free-standing
  aupd fine wires}.
\newblock \emph{\bibinfo{journal}{Solid State Commun.}}
  \textbf{\bibinfo{volume}{57}}, \bibinfo{pages}{411 -- 416}
  (\bibinfo{year}{1986}).

\bibitem{Wellstood1994}
\bibinfo{author}{Wellstood, F.~C.}, \bibinfo{author}{Urbina, C.} \&
  \bibinfo{author}{Clarke, J.}
\newblock \bibinfo{title}{Hot-electron effects in metals}.
\newblock \emph{\bibinfo{journal}{Phys. Rev. B}} \textbf{\bibinfo{volume}{49}},
  \bibinfo{pages}{5942--5955} (\bibinfo{year}{1994}).

\bibitem{Rusby2002}
\bibinfo{author}{Rusby, R.~L.} \emph{et~al.}
\newblock \bibinfo{title}{The provisional low temperature scale from 0.9 mK to
  1 K, plts-2000}.
\newblock \emph{\bibinfo{journal}{J. Low Temp. Phys.}}
  \textbf{\bibinfo{volume}{126}}, \bibinfo{pages}{633--642}
  (\bibinfo{year}{2002}).

\bibitem{Clark2010}
\bibinfo{author}{Clark, A.~C.}, \bibinfo{author}{Schwarzwälder, K.~K.},
  \bibinfo{author}{Bandi, T.}, \bibinfo{author}{Maradan, D.} \&
  \bibinfo{author}{Zumbühl, D.~M.}
\newblock \bibinfo{title}{Method for cooling nanostructures to microkelvin
  temperatures}.
\newblock \emph{\bibinfo{journal}{Rev. Sci. Instrum.}}
  \textbf{\bibinfo{volume}{81}} (\bibinfo{year}{2010}).

\bibitem{Casparis2012}
\bibinfo{author}{Casparis, L.} \emph{et~al.}
\newblock \bibinfo{title}{Metallic coulomb blockade thermometry down to 10 mK
  and below}.
\newblock \emph{\bibinfo{journal}{Rev. Sci. Instrum.}}
  \textbf{\bibinfo{volume}{83}} (\bibinfo{year}{2012}).

\end{thebibliography}

\vspace{\baselineskip}

{\large\noindent\textbf{Acknowledgments}}\\
{\footnotesize This work was supported by the European Research Council (ERC-2010-StG-20091028, no. 259033), the French RENATECH network, the national French program `Investissements d'Avenir' (Labex NanoSaclay, ANR-10-LABX-0035) and the European Seventh Framework Program (EU FP7, no. 263455).

\vspace{\baselineskip}
{\large\noindent\textbf{Author contributions}}\\
{\footnotesize Measurements and analysis: F.P. with inputs from Z.I.; low temperature setup: F.P. with inputs from A.A. and S.J.; noise measurement setup: F.D.P, S.J., A.A. and F.P.; HEMTs nanofabrication: Y.J.; heterojunction growth: A.C., A.O. and U.G.; sample nanofabrication: F.D.P and A.A; manuscript preparation: F.P. with inputs from A.A. and U.G.; project planning and supervision: F.P.}

\vspace{\baselineskip}
{\large\noindent\textbf{Competing Financial Interests}}\\
{\footnotesize The authors declare no competing financial interests.}

\vspace{\baselineskip}
{\large\noindent\textbf{Additional information}}\\
{\footnotesize\noindent\textbf{Supplementary information} accompanies this paper.}


\end{document}





\section{Supplementary Figures}

\begin{figure}[h]
\renewcommand{\figurename}{\textbf{Supplementary Figure}}
\centering\includegraphics[width=0.7\columnwidth]{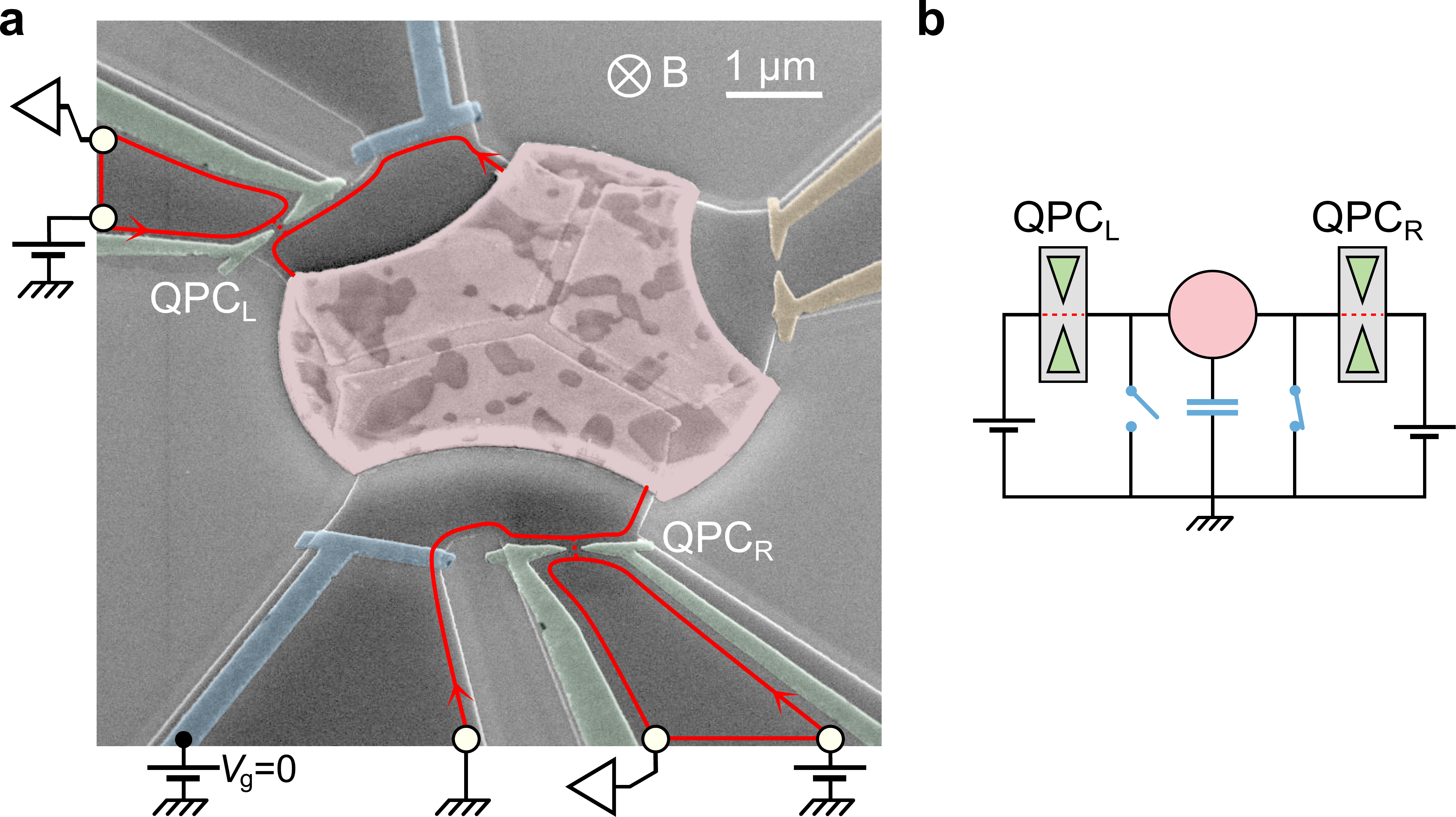}
\caption{\textbf{Quantum shot-noise configuration}.
(\textbf{a}) Colored micrograph of the measured device.
The displayed edge currents (red lines) here correspond to the circuit configuration used for the quantum shot noise measurements across QPC$_R$.
The right short-circuit switch is closed by applying $V_\mathrm{g}=0$.
(\textbf{b}) Corresponding circuit schematic.
}
\label{fig-SIfig1}
\end{figure}

\begin{figure}[hb]
\renewcommand{\figurename}{\textbf{Supplementary Figure}}
\centering\includegraphics[width=0.85\columnwidth]{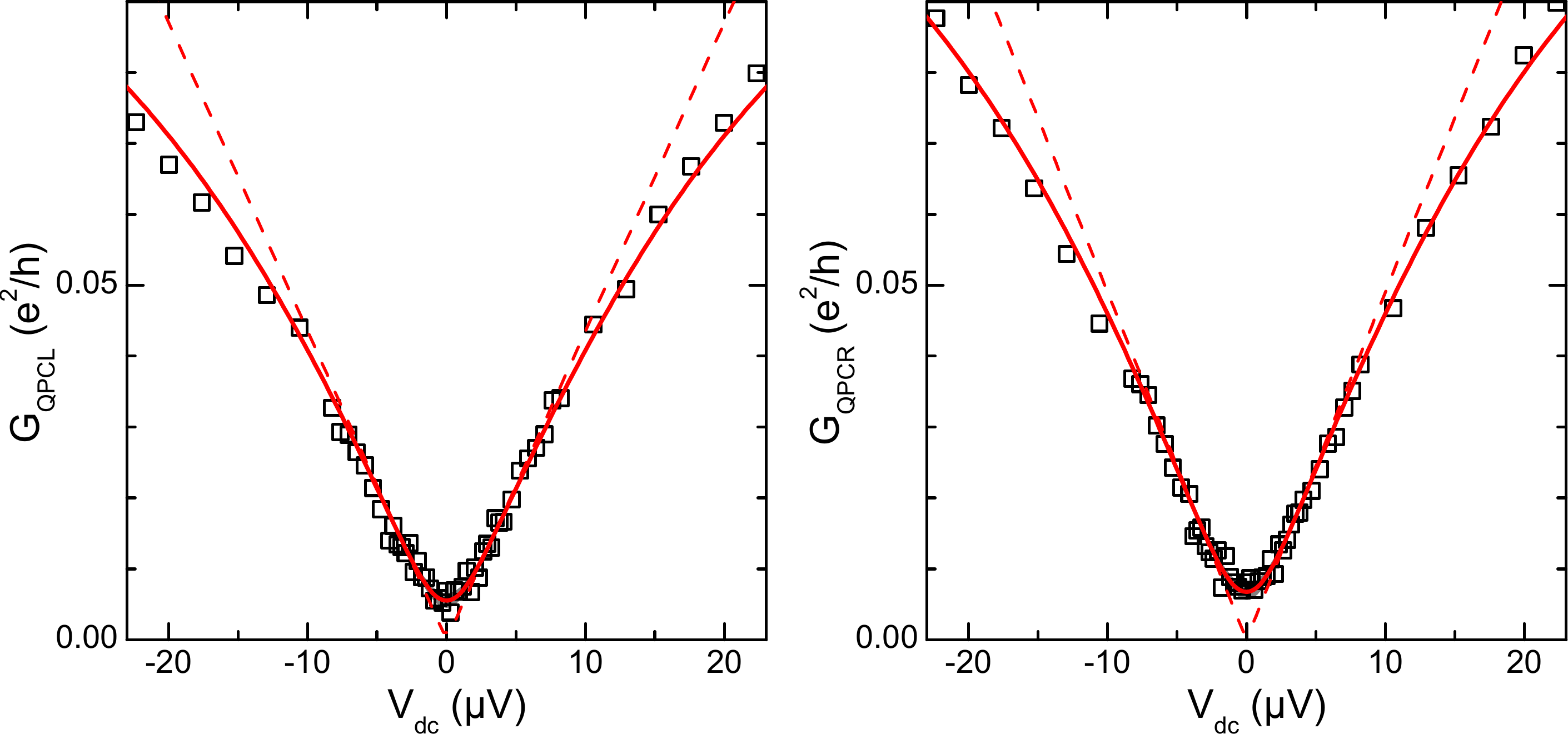}
\caption{\textbf{Dynamical Coulomb blockade data vs predictions}.
The same data (symbols) and theoretical predictions (continuous lines: numerical calculations of the full prediction; dashed lines: $T=0$ predictions at $eV_\mathrm{dc}\ll E_C$ of Eq.~4) shown in Fig.~2c are here displayed up to larger dc bias voltages.
}
\label{fig-SIfig2}
\end{figure}

\begin{figure}[hb]
\renewcommand{\figurename}{\textbf{Supplementary Figure}}
\centering\includegraphics[width=0.85\columnwidth]{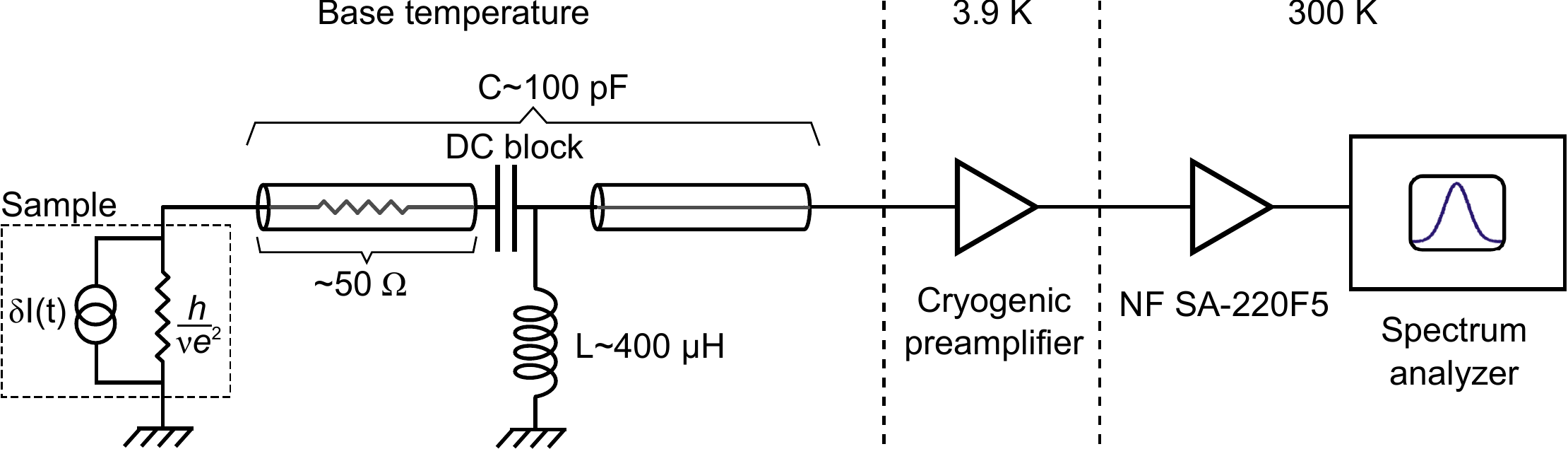}
\caption{\textbf{Noise measurement setup}.
The current fluctuations $\delta I(t)$ are converted into voltage fluctuations through the resonator impedance $Z$ consisting on the on-chip quantum Hall resistance $R=h/\nu e^2$, with $\nu$ the filling factor ($\nu=6$ for the magnetic field $B=1.4$~T) in parallel with a LC tank. The latter consists of a superconducting inductance $L\sim400~\mu$H and the capacitance $C\sim100~$pF developing along the coaxial lines resulting in a resonant frequency of $f_\mathrm{LC}\simeq0.84$~MHz. The bandwidth at $-3~\mathrm{dB}$ of the overall resonator $Z$ is equal to $1/2\pi RC$ and ranges from $370$ to $120~\mathrm{kHz}$ for $\nu=6$ to $2$. The achieved relative precision of the voltage fluctuation measurement can be increased by two parameters: the integration time $t_i$ and for a white noise, the frequency interval $\Delta f$ on which the measurement is performed. Indeed, the relative precision scales as $1/\sqrt{N}$ with $N$ the number of measured samples, where $N$ is proportional to the product of $t_i$ and $\Delta f$. In practice, at $\nu=6$ and $T=6~\mathrm{mK}$, we used $t_i=40~\mathrm{s}$ per point and the frequency window $\Delta f=\big[0.68, 1\big]~\mathrm{MHz}$ that roughly matches the resonator bandwidth. Moreover, note that the data shown in Fig. 2a are an average of $131$ sweeps, each consisting of $31$ points, and were measured in about $2$ days. The electronic temperature can be extracted directly, without calibration of the noise measurement setup, from the integrated raw signal.
}
\label{fig-SIfig3}
\end{figure}

\clearpage
\section{Supplementary Notes}

{\large\noindent\textbf{Supplementary Note 1: Low temperature components of the experimental setup}}\\


{\noindent\textbf{Sample installation.}}

The measured sample is glued to the grounded gold back-plane of a ceramic leadless chip carrier (Kyocera, part number: PB-44713) and electrically connected by aluminum wire bonding.

The ceramic chip carrier is then plugged in a plastic socket (E-Tec, part number: LCC-044-H210-55) inside the top-loaded sample holder.
The socket, reinforced laterally with epoxy resin (Stycast 2850FT with catalyst 9) and at the bottom with a stainless steel plate, is permanently screwed to the top-loaded sample holder.

The inner stainless steel shield, whose bottom inside surface is covered by a thin layer of microwave absorber (Eccosorb CR-124 epoxy resin), and the outer gold plated brass shield are screwed into position (see Fig.~4b).

The sample holder is then inserted from the top of our dilution refrigerator (Triton 200 from Oxford instruments), through rotating radiation shields, and screwed to the mixing chamber (Triton top loading option).

\vspace{\baselineskip}
{\noindent\textbf{Measurement lines filtering and thermalization at low temperature.}}

The high-frequency filtering and initial thermalization is done following Ref.~20, using resistive microcoaxial cables.
For each measurement line, one meter of a narrow resistive NiCrAlSi (Isaohm) wire (diameter 75~$\mu$m, resistance $300~\Omega/$m) is inserted into a CuNi tube of inner diameter 260~$\mu$m.
The CuNi tubes are tightly coiled on a copper plate screwed to the mixing chamber of the dilution refrigerator.

The low frequency filtering is performed inside the top-loaded sample holder, using simple $RC$-filters with CMS components (nichrome resistances from Vishay TNPW series, C0G ceramic capacitors from Murata GRM series).

Additional high-frequency filtering is provided by $\sim30$~cm long CuNi microcoaxes (Coax Co. Ltd., part number: SC-040/50-CN-CN) between $RC$-filters and the inner stainless steel shield.

Inside the inner stainless steel shield, the strongest thermal anchoring of each measurement line is realized by dipping $\sim5$~cm of a copper wire coated with a thin insulating layer into a conductive silver epoxy (Epotek, part number: H20E) together with a thermalized copper braid.

\clearpage
{\large\noindent\textbf{Supplementary Note 2: Noise measurement setup.}}\\

The current fluctuations measurements are performed using a setup very similar to that described in more details in the supplementary material of Ref.~11, here implemented in a cryogen-free dilution refrigerator.

A schematic representation of the amplification chain for the current fluctuations measurement is shown in Supplementary Fig.~3. 
The current fluctuations are converted on-chip into voltage fluctuations using the well-defined quantum Hall resistance $R=h/\nu e^2$, with $\nu$ the filling factor ($\nu=6$ at $B\simeq1.4$~T).
The most crucial element of the amplification chain is the home-made cryogenic voltage preamplifier (see Refs.~[11,32,33]), which is thermalized to the $3.9$~K plate.
It is operated slightly below 1~MHz, where it shows the best performances and where the electrical noise induced by the pulse tube vibrations is found negligible.
For this purpose, we shift the experimental frequency bandwidth with a parallel $L$--$C$ tank circuit of resonant frequency $f_\mathrm{res}\simeq0.84$~MHz. 
The capacitance $C\sim100~$pF is the capacitance that develops along the coaxial lines connecting the sample to the cryogenic preamplifier, whereas the inductance $L\sim400~\mu$H is realized with a superconducting coil thermally anchored at base temperature but located away from the magnetic field.
The current fluctuations signal remains within the same frequency bandwidth $\sim1/2\pi RC$, now around $f_\mathrm{res}$.
Note that we use a dc block to ascertain that the amplification chain is not influenced by the dc voltage bias across the sample.

The spectral density of the measured current fluctuations is integrated over a frequency window that is optimized separately for each value of $\nu$.
At $\nu=6$ ($B\simeq1.4$~T), where we obtain the lowest electronic temperature $T\simeq6.0$~mK, the optimized integration window is $f\in[0.68,1]~$MHz.
As described below, this raw integrated signal $S_I^\mathrm{raw}$ versus the dc bias voltage $V_\mathrm{dc}$ applied to the QPC can be used directly to extract the electronic temperature, without calibration of the noise measurement setup and without the knowledge of the transmission probabilities $\{\tau_n\}$ across the QPC.

More precisely, the excess raw integrated signal is simply proportional to the excess spectral density $\Delta S_I^\mathrm{raw}\equiv S_I^\mathrm{raw}-S_I^\mathrm{raw}(V_\mathrm{dc}=0)=G\Delta S_I$, with $G$ the effective amplification chain gain (depending on the frequency window). 
The excess raw integrated signal can then be fitted using Eq.~1 with two free parameters: the electronic temperature $T$ and the product $G\sum\tau_n(1-\tau_n)$.
The product $G\sum\tau_n(1-\tau_n)$ is given by the temperature-independent linear slope of $\Delta S_I^\mathrm{raw}(V_\mathrm{dc})$ at large $|V_\mathrm{dc}|\gg k_BT/e$, and the temperature $T$ is obtained from the zero to large voltage bias crossover.
In practice, we start directly with the full (not excess) raw integrated signal and therefore also use the zero bias offset $S_I^\mathrm{raw}(V_\mathrm{dc}=0)$ as a fit parameter.

With the additional knowledge of the transmission probabilities $\{\tau_n\}$ across the QPC, the effective amplification chain gain $G$ can be extracted from the fit parameter $G\sum\tau_n(1-\tau_n)$.
Although not necessary to extract the temperature $T$, such a calibration is used in the top panel of Fig.~2a in order to display the measured excess noise $\Delta S_I$ in units of A$^2$/Hz.
Note that in practice the QPC is set to a single half transmitted channel, whose precise transmission probability $\tau\simeq0.55$ is measured simultaneously to the current fluctuations.
